\documentclass[floatfix,showpacs,superscriptaddress,nofootinbib,twocolumn,showkeys,prl]{revtex4-2}

\usepackage[utf8]{inputenc}

\usepackage{graphicx}
\usepackage{xcolor}
\usepackage{amsmath}
\usepackage{hyperref}

\usepackage{amssymb}
\usepackage{multirow,array}
\usepackage{lipsum}
\usepackage{enumitem}

\newcommand{\sqrts}{\sqrt{s_\textrm{NN}}}

\newcommand{\prlsection}[1]{\noindent \textbf{\emph{#1}} -- }

\begin{document}

\title{
Early-times Yang-Mills dynamics and the characterization of strongly interacting matter with statistical learning
}

\author{Matthew Heffernan}
\affiliation{Department of Physics, McGill University, Montr\'{e}al QC H3A\,2T8, Canada}

\author{Charles Gale}
\affiliation{Department of Physics, McGill University, Montr\'{e}al QC H3A\,2T8, Canada}
\author{Sangyong Jeon}
\affiliation{Department of Physics, McGill University, Montr\'{e}al QC H3A\,2T8, Canada}

\author{Jean-Fran\c{c}ois Paquet}
\affiliation{Department of Physics and Astronomy, Vanderbilt University, Nashville TN 37235}
\affiliation{Department of Mathematics, Vanderbilt University, Nashville TN 37235}

\date{\today}

\begin{abstract}
In ultrarelativistic heavy-ion collisions, a plasma of deconfined quarks and gluons is formed within $1$~fm/c of the nuclei's impact. The complex dynamics of the collision before $\approx 1$~fm/c is often described with parametric models,
which affect the predictivity of calculations.
In this work, we perform a systematic analysis of LHC measurements from Pb-Pb collisions, by combining an \emph{ab-initio} model of the early stage of the collisions with a hydrodynamic model of the plasma. We obtain state-of-the-art constraints on the shear and bulk viscosity of quark-gluon plasma.
We mitigate the additional cost of the ab-initio initial conditions by combining Bayesian model averaging with transfer learning, allowing us to account for important theoretical uncertainties in the hydrodynamics-to-hadron transition.
We show that, despite the apparent strong constraints on the shear viscosity, 
metrics that balance the model's predictivity with its degree of agreement with data do not prefer a temperature-dependent specific shear viscosity over a constant value.
We validate the model by comparing with discriminating observables not used in the calibration, finding excellent agreement.
\end{abstract}

\maketitle

\prlsection{Introduction}

The fundamental degrees of freedom of Quantum Chromodynamics (QCD) -- the theory of the nuclear strong interaction -- are quarks and gluons. 
In the overlap region of two heavy nuclei colliding at ultrarelativistic energies, a dense plasma of such quarks and gluons is formed \cite{[{See, for example, }][{, and references therein.}]Busza:2018rrf}. The rapid expansion of this strongly-interacting plasma can be described with relativistic viscous hydrodynamic \cite{Gale:2013da,Heinz:2013th,DerradideSouza:2015kpt,Romatschke:2017ejr}, providing a window into the transport properties of hot and dense nuclear matter. This theoretical development  brought the heavy-ion program into a characterization phase. 

Early analyses of measurements at the Relativistic Heavy Ion Collider established that the specific shear viscosity $\eta/s$ (the coefficient of shear viscosity over the entropy density)  of quark-gluon plasma must be small, of the order of $1/4\pi$ \cite{Romatschke:2007mq}. 
Since these early analyses, considerable effort has been devoted to quantify the full temperature dependence of both shear and bulk viscosity \cite{Gale:2013da}. This is a formidable undertaking, since the quark-gluon plasma is formed in heavy-ion collisions for less than $10^{-23}$ seconds (10 Fermi/$c$), evolving through multiple distinct stages between the impact of the nuclei and the final measurement of particle showers by the detectors. Furthermore, it was understood early-on that shortcomings in the theoretical understanding of any stage of the collision translate into uncertainties in the determination of the viscosities~\cite{Luzum:2008cw}. An important development has been the use of Bayesian methods to quantify precisely the uncertainties on the viscosities extracted from model-to-data comparisons. A number of Bayesian analyses have now been performed~\cite{Petersen:2010zt,Novak:2013bqa,Sangaline:2015isa,Bernhard:2015hxa,Bernhard:2016tnd,Moreland:2018gsh,Bernhard:2019bmu,JETSCAPE:2020mzn,Nijs:2020ors,Nijs:2020roc,Nijs:2021clz,Parkkila:2021tqq,Nijs:2023yab}, comparing modern multistage models of heavy-ion collisions with broad sets of measurements from the Relativistic Heavy Ion Collider and the Large Hadron Collider.

A long-standing source of uncertainty in heavy-ion collisions is the early stage of the collision, which provides the initial conditions for the relativistic viscous fluid dynamic simulation. While multiple early-stage models have been studied and compared in the literature, current Bayesian analyses have mostly relied on a flexible parametric model of the impact of the collision \cite{Moreland:2014oya}, sometimes augmented with more parameters~\cite{Nijs:2023yab} or supplemented with simplified dynamical models of the early stage of the collision~\cite{JETSCAPE:2020mzn,Nijs:2020roc,Nijs:2023yab}. In this work, we replace this parametric approach with an ab-initio model of the early stage of the collision, the IP-Glasma model \cite{Schenke:2012hg,Schenke:2012wb}. 
The foundation of the IP-Glasma model is to describe the nuclei, before their collision, as two color glass condensates, which is a limiting representation of the low-energy gluons in large nuclei accelerated at high energies~\cite{McLerran:1993ni}. The collision of the two color glass condensates generates gluon fields~\cite{Bartels:2002cj,Kowalski:2003hm} which are evolved in time with the classical Yang-Mills equations. This model is known for its success in describing a wide range of heavy-ion measurements with a limited set of parameters~\cite{Gale:2012rq,Schenke:2012wb,Gale:2013da,Ryu:2015vwa,McDonald:2016vlt,Ryu:2017qzn,Schenke:2020mbo,Gale:2021emg}.

Using these ab-initio IP-Glasma initial conditions, we show in this work that our state-of-the-art model of heavy-ion collisions can describe a wide set of heavy-ion measurements. We employ Bayesian techniques to constrain the parameters in our approach, including shear and bulk specific viscosity coefficients. We also illustrate the power of transfer learning to account for uncertainty from modeling choices in simulations of heavy-ion collisions. This paper is a companion paper to a larger write-up \cite{ThesisPRC}  that contains many of the details omitted here for the sake of brevity. 

\prlsection{Model-data comparison}
We consider a broad range of measurements from Pb-Pb collisions with $\sqrts = 2.76$ TeV center-of-mass energy per nucleon pair at the Large Hadron Collider \cite{ThesisPRC}: 
(i) the number of charged hadrons per unit pseudorapidity $dN_{ch}/d\eta$ \cite{ALICE:2010mlf}; 
(ii) the number of identified charged hadrons per unit rapidity $dN_i/dy$, $i \in \{\pi, p, k\}$  \cite{ALICE:2013mez}; 
(iii) the transverse energy per unit pseudorapidity $dE_T/d\eta$ \cite{ALICE:2016igk}; 
(iv) the mean transverse momenta of identified hadrons $\langle p_T \rangle_i$, $i \in \{\pi, K, p\}$ \cite{ALICE:2013mez}; 
(v) transverse momentum fluctuations $\delta p_T/\langle p_T \rangle$ \cite{ALICE:2014gvd};
(vi) two-particle radial Fourier coefficients \cite{ALICE:2011ab}; 
(vii) four-particle transverse Fourier coefficient $v_2\{4\}$ \cite{ALICE:2016ccg};
(viii) event plane correlators \cite{ATLAS:2014ndd, Acharya:2017zfg}; 
(ix) nonlinear response coefficients that quantify mixing between higher and lower-order modes $\chi_{n,mk}$ \cite{Acharya:2017zfg}; 
(x) linear and non-linear flow modes \cite{Acharya:2017zfg}. 
These observables represent a diverse set of the available measurements that can be calculated with reasonable statistical accuracy.
Additional observables are used to verify the model's predictivity, as discussed below.

The Pb-Pb collisions are simulated in three distinct phases. The ab-initio IP-Glasma model describes the first stage of the collision, from the nuclei's impact to the resulting production of color fields and their subsequent initial evolution. The energy-momentum tensor of these gluons is then used as initialization for relativistic viscous hydrodynamics~\cite{Schenke:2010nt,Schenke:2010rr,Paquet:2015lta}. Because we focus on measurements made at midrapidity, we assume boost invariance along the collision axis~\cite{Bjorken:1982qr}. An equation of state based on to lattice quantum chromodynamics calculations is used in the hydrodynamic phase~\cite{HotQCD:2014kol, Bernhard:2018hnz,eosmaker}. The viscosities of the plasma are parametrized with the same flexible parametrizations as used in Ref.~
\cite{JETSCAPE:2020mzn,SIMSPRL}.\footnote{For the specific shear viscosity, we use 
a piecewise linear function, with low temperature slope and a high temperature slope separated by an single inflection point; the inflection point is often expected to be a minimum, as observed in the temperature dependence of other fluids~\cite{Csernai:2006zz}, although we allow it to be either a minimum or a maximum. The specific bulk viscosity is parametrized with a skewed Cauchy distribution, allowing it to peak in the deconfinement region, and decrease at different rate above and below this peak. All details are in Ref. \cite{ThesisPRC}.}
 Finally, at a chosen local energy density, the energy-momentum tensor of the fluid is matched to hadronic momentum distributions, and relativistic hadronic transport~\cite{smash} is used for the third phase of the collision. There is uncertainty in determining the hadronic momentum distributions at this transition, and in consequence two different models are studied: one based on Grad's 14-moment approximation of the Boltzmann equation\cite{Grad,Israel:1976tn, Israel:1979wp, Teaney:2003kp, Dusling:2009df, Monnai:2009ad, Dusling:2011fd, Denicol:2012cn}, and the other using a Chapman-Enskog expansion in the relaxation time approximation\cite{chapman1990mathematical, 
    ANDERSON1974466, Jaiswal:2014isa}.

Two parameters are associated with the initial IP-Glasma phase: the first relates the scale at which gluon saturation occurs to the density of color charges, and the second is the hydrodynamic initialization time. Eight more parameters are used to describe the temperature dependence of the viscosities (see previous footnote), and one final parameter is used to define the energy density for the transition between hydrodynamics and hadronic transport. The smaller number of parameters in the first stage of the collision, at least four fewer parameters than previous analysis, leads to a total of $11$~parameters. We will discuss later that the number of parameters can be further reduced to $8$ by curtailing the flexibility of the shear viscosity parametrization.

Model-to-data comparison is performed with Bayesian inference. To access the model's predictions for different value of viscosities and other parameters, 
we build a fast surrogate using Gaussian process emulators. We define the admissible range of the model parameters using Generalized Normal priors, and we sample 350 sets of parameters with a Maximum Projection Latin hypercube~\cite{MaxPro}. Model simulations are performed for all 350 parameter sets. Rather than training a surrogate model for each observable of interest, a dimensional reduction using principal component analysis is performed on the observables. We then train Gaussian processes on 20 principal components, which capture 90.6\% of the models information, and use these to emulate the model's output for all observables.

The above emulation is performed on calculations that use the Grad's 14-moment model to describe the transition between hydrodynamics and hadronic transport. Because of the uncertainty in this transition, it is important to use more than one model to compare with measurements. Repeating the entire procedure for our second model (Chapman-Enskog) would have been inefficient, however, given that only a single part of the model is changed. This is a general challenge in studying theoretical uncertainties in heavy-ion collision modelling: it is a high priority for the community to quantify model uncertainties, but the computational costs are often prohibitive.
We address this problem with transfer learning, building upon Ref.~\cite{Liyanage:2022byj}. We write the model's output as
\begin{equation}
	f_{\textrm{Chapman-Enskog}}(x) = \rho f_{\textrm{Grad}}(x) + \delta (x)
\end{equation} 
where $\rho$  and $\delta$ are respectively a linear correlation term and a discrepancy term between the source and target models. Constraints on $\rho$ and $\delta (x)$ are obtained by comparing with 50 design points evaluated with the Chapman-Enskog model. 
This technique 
allows us to estimate, at a substantially reduced computational cost, the uncertainty in the transition from hydrodynamics to individual hadrons. We emphasize the potential of this approach to study a wider range of theoretical uncertainties in the study of heavy-ion collisions. Additional details of the implementation of transfer learning and emulation can be found in Ref.~\cite{Heffernan:2023utr}.

We use a Gaussian likelihood function to compare models and measurements~\cite{Heffernan:2023utr}. We evaluate the posterior $p(H|d,I)$, given by the product of the prior $p(H,I)$ and the likelihood $p(d|H,I)$, normalized by the Bayes evidence $p(d,I)$:
\begin{equation}
	p(H|d,I) = \frac{p(H,I) p(d|H,I)}{p(d,I)}. \label{eq:bayes}
\end{equation}

\prlsection{Constraints on shear and bulk viscosity}

\begin{figure*}[tbh]
    \centering
    \includegraphics[width=0.75\textwidth]{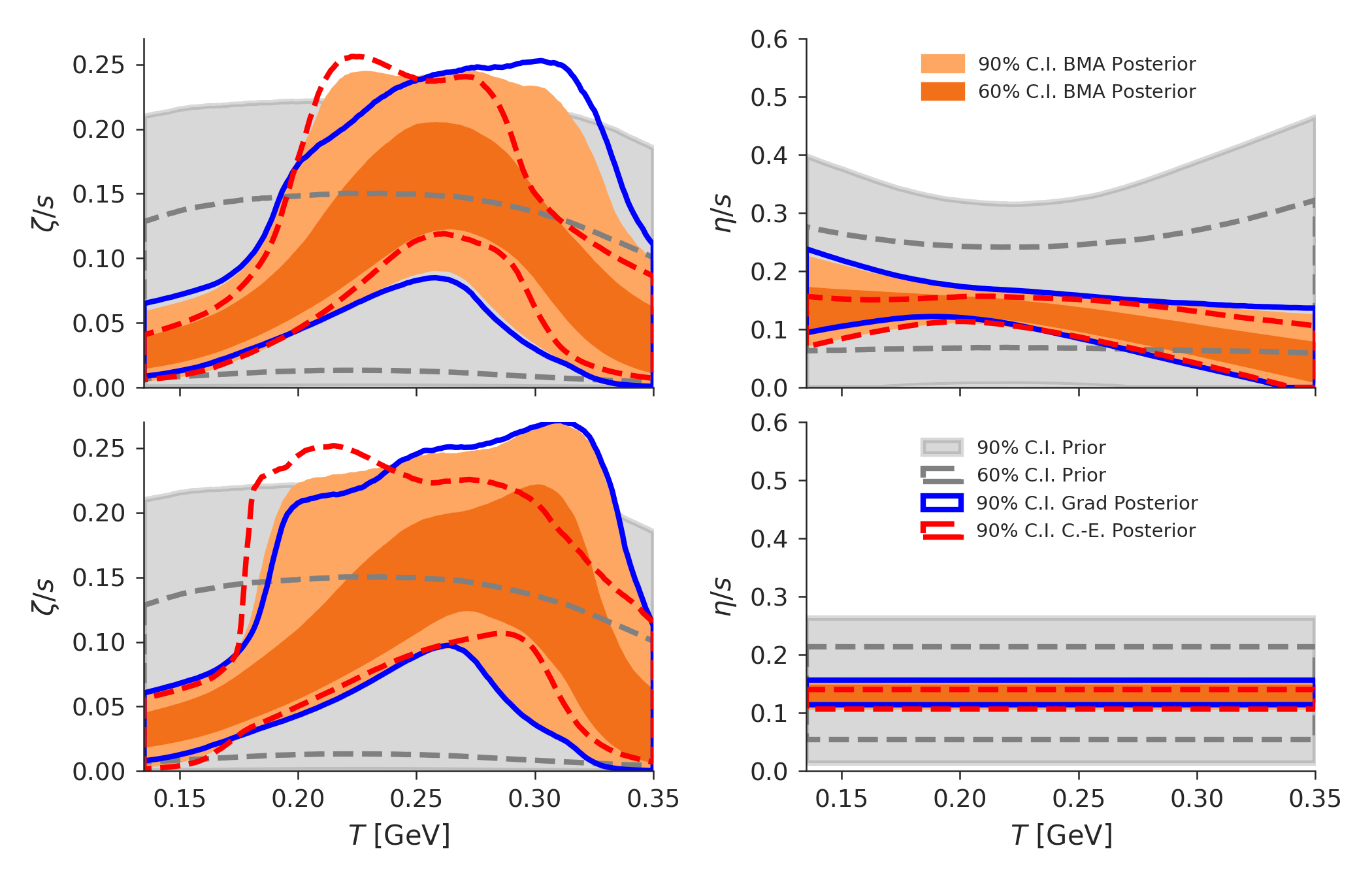}
    \caption{Bayes Model Averaged viscous posterior with shear viscosity allowed to vary with temperature (top panels) and the Bayes Model Averaged viscous posterior when the shear viscosity is constant with temperature (bottom panels). Individual 90\% credible intervals for the Grad and Chapman-Enskog viscous corrections that underlie the averaging are shown in blue and red, respectively.}
    \label{fig:full-study-BMA-viscous-posterior}
\end{figure*}

The 90\% credibility intervals for the shear and bulk viscosities, obtained from the Bayesian calibration of our multistage model of heavy-ion collisions to the LHC data sets listed in the previous section, are shown at the top of Figure~\ref{fig:full-study-BMA-viscous-posterior}. 
Results are shown for both the Grad and the Chapman-Enskog model of the hydrodynamics-to-hadrons transition. The results are similar for both models. At low temperature ($T \approx 150$~MeV), large values of the specific bulk viscosity $\zeta/s$ are disfavored; conversely, significant values of bulk viscosity are favored between temperatures of 200 and 300 MeV. For the specific shear viscosity $\eta/s$, the model-to-data comparison leads to intermediate values $\approx~0.1$ to $0.2$ with a modest temperature dependence. 
To obtain a single constraints from multiple models, we use Bayesian model averaging~\cite{hoeting1999bayesian,SIMSPRL,Heffernan:2023utr}, which yields the orange (dark) band shown in Figure~\ref{fig:full-study-BMA-viscous-posterior}.  

The weak temperature dependence of $\eta/s$ found in Figure~\ref{fig:full-study-BMA-viscous-posterior} suggests that good agreement with measurements could also be obtained using a constant value of the specific shear viscosity. 
This question goes beyond temperature dependence of $\eta/s$ and it directly related to the predictivity of the model. 
A model with a temperature-dependent shear viscosity will always be able to describe measurements as well or better than a model with a constant $\eta/s$ (under the straightforward assumption that the prior of the temperature-dependent $\eta/s$ includes the constant values). The more important question is: given the additional model flexibility from the temperature-dependent $\eta/s$, is there a statistically meaningful difference in the model's agreement with data? Metrics that balance model predictivity and its degree of agreement with data are well known (see e.g. \cite{Liddle:2004nh,Liddle:2007fy}). In this work, we use the Bayes factor, as in Ref.~\cite{JETSCAPE:2020mzn}. This is calculated by taking the ratio of the Bayes evidences -- the denominator $p(d,I)$ in Bayes' theorem Eq.~\ref{eq:bayes} -- between two posterior distributions.

The Bayes factor between constant and temperature-dependent $\eta/s$ is $\ln B = 0.2 \pm 2.4$ showing a negligible preference for temperature-dependent $\eta/s$. This is the result with the Grad hydrodynamics-to-hadron model; the result is very similar with the Chapman-Enskog RTA model, at $\ln B = 1 \pm 4$. 
We note that Ref.~\cite{JETSCAPE:2020mzn} had also found similar Bayes factors with a constant and temperature-dependent $\eta/s$.

While the degree of constraint on $\eta/s$ is significant, conclusive evidence for temperature dependence remains elusive.  In essence, the additional complexity of the model with a temperature-dependence $\eta/s$ is not justified by the corresponding small improvement in describing experimental results. The posterior for the viscosities, assuming a constant $\eta/s$, are shown in the bottom panels of Figure~\ref{fig:full-study-BMA-viscous-posterior}. Constraints on the bulk viscosity are similar for both the constant and the temperature-dependent $\eta/$.

\begin{figure}[htb]
      	\begin{centering}
      		\includegraphics[width=\columnwidth]{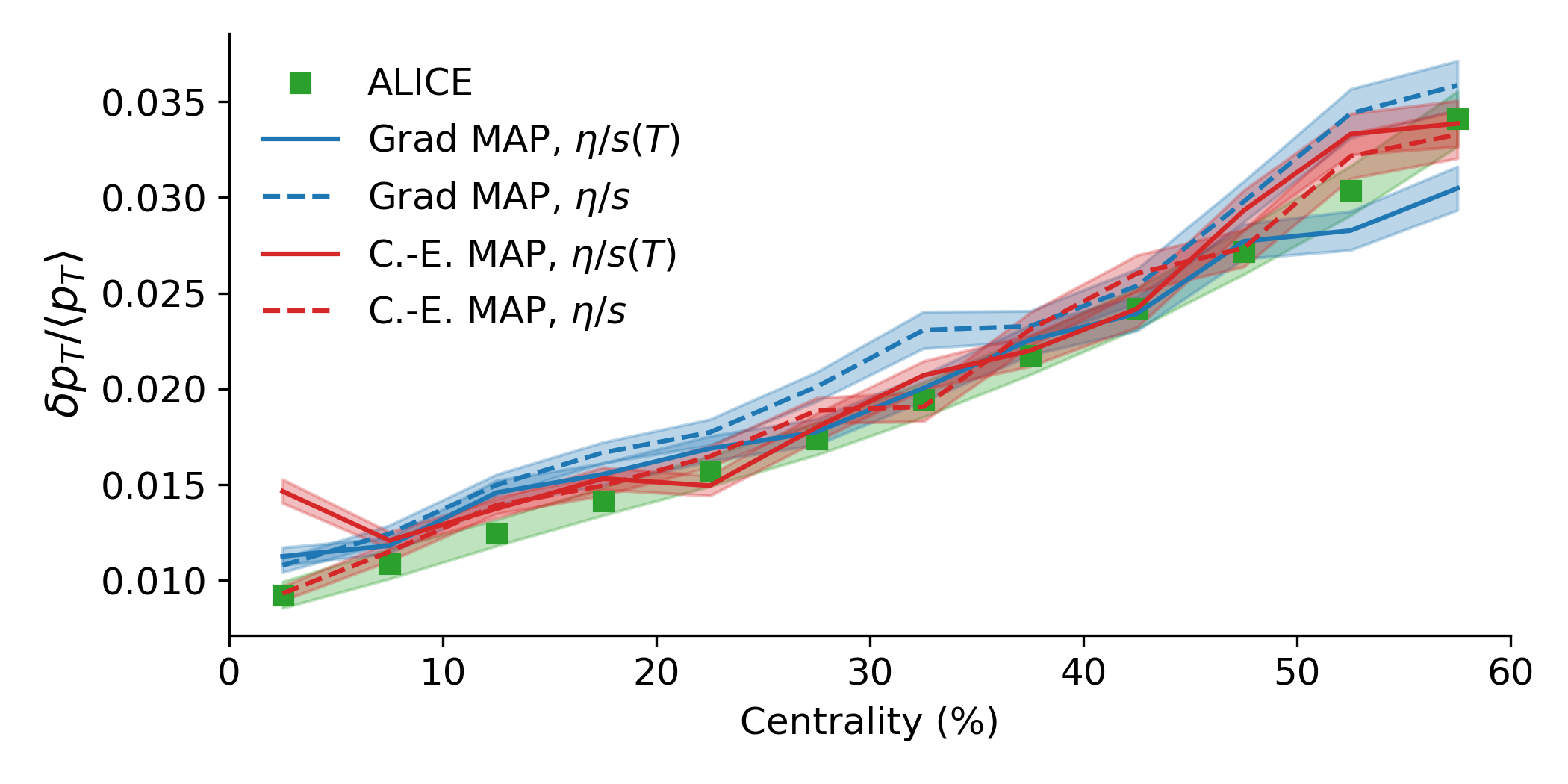}
      		\caption{Postdiction of $\delta p_T / \langle p_T \rangle$ at Maximum a Posteriori. Data from \cite{ALICE:2014gvd}.}
      		\label{fig:full-study-MAP-pTfluct}
      	\end{centering}
\end{figure}

Confirming the results from the Bayes factor, we find a similar agreement with measurements for both the constant and temperature-dependent value of $\eta/s$ shown in Figure~\ref{fig:full-study-BMA-viscous-posterior}. 
We show in Figure~\ref{fig:full-study-MAP-pTfluct} the results for one of the measurements used for the calibration, the transverse momentum fluctuation. The results for a constant and a temperature-dependent $\eta/s$ are consistent within statistical uncertainties. We verified that the same conclusion can be reached for all measurements used in the calibration (see Ref.~\cite{Heffernan:2023utr}). 

\begin{figure}[!htb]
      	\begin{centering}
      		\includegraphics[width=\columnwidth]{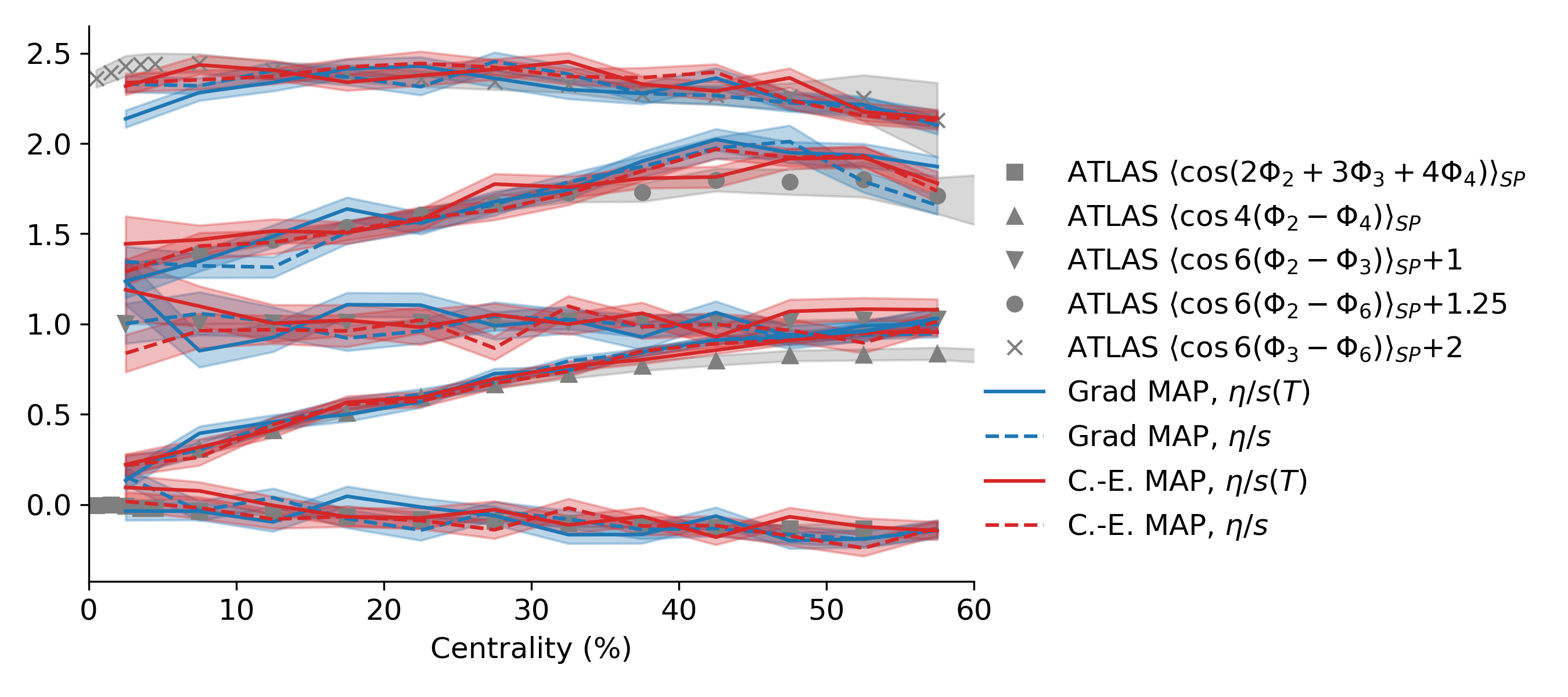}
      		\caption{Prediction of ATLAS event-plane correlations at Maximum a Posteriori. Data~\cite{ATLAS:2014ndd}  and calculations are shifted for clarity.}
      		\label{fig:full-study-MAP-predictions-correlators-ATLAS}
      	\end{centering}
\end{figure}

We now turn to comparison with measurements that were not used to constrain the viscosities and other model parameters. We first compare with event-plane correlations, which describe correlation between the symmetry axes of different Fourier harmonics \cite{ATLAS:2014ndd, Acharya:2017zfg}. As can be seen in Figure~\ref{fig:full-study-MAP-predictions-correlators-ATLAS}, we find good agreement with four different event-plane correlators, with either model of hydrodynamics-to-hadrons transition and with either a constant $\eta/s$ and a temperature-dependent one.
\begin{figure}[htb]
      	\begin{centering}
      		\includegraphics[width=0.49\textwidth]{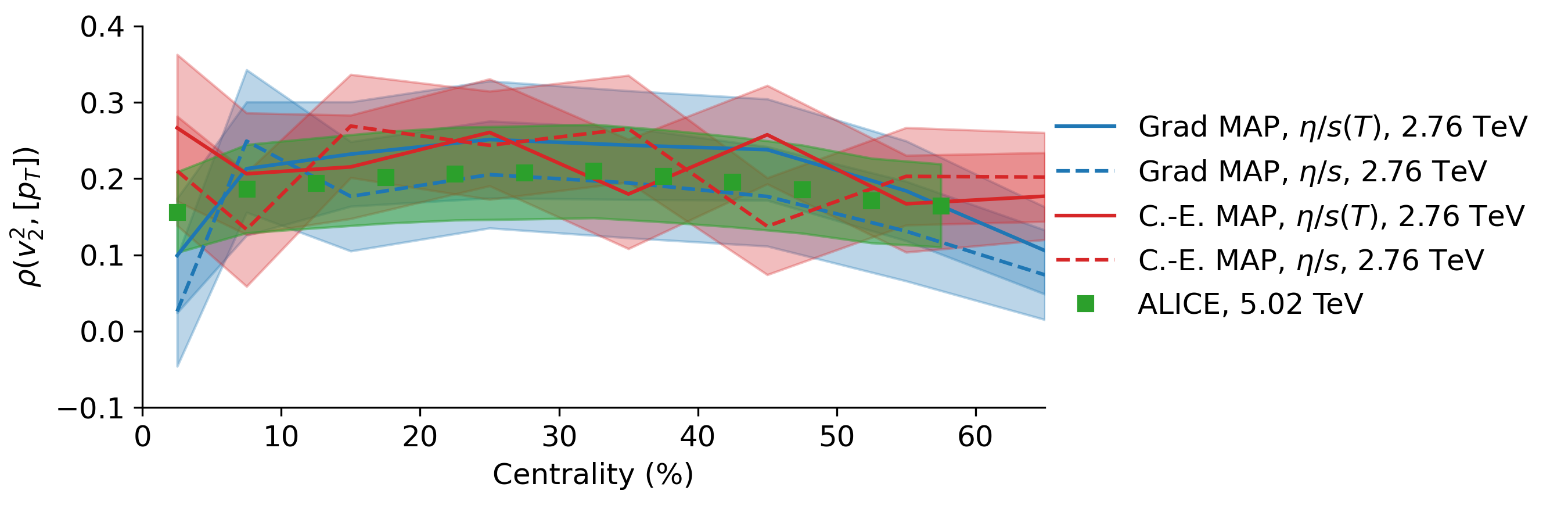}
      		\caption{Correlation between $v_2^2$ and $p_T$ at Maximum a Posteriori, compared to data from a higher-energy collision. Note that data~\cite{ALICE:2021gxt} are at $\sqrt{s_{NN}}=5.02$ TeV while the MAP predictions are at $\sqrt{s_{NN}}=2.76$ TeV.}
      		\label{fig:full-study-MAP-predictions-rho2}
      	\end{centering}
\end{figure}

We further make a prediction for the correlation between momentum anisotropy fluctuations and average transverse momentum fluctuations. This observable was measurements for $\sqrts=5.02$~TeV Pb-Pb collisions by the ALICE~\cite{ALICE:2021gxt} and ATLAS~\cite{ATLAS:2022dov} collaborations. We make a prediction for $\sqrts=2.76$~TeV Pb-Pb collisions, which we superimpose with the $\sqrts=5.02$~TeV Pb-Pb collisions by the ALICE~\cite{ALICE:2021gxt}. If, as expected, this observable is measured to have a very small $\sqrts$ dependence, we expect excellent agreement of our prediction with the measurement.

\prlsection{Summary}

We compared a multistage model of ultrarelativistic nuclear collisions with a large set of measurements from the Large Hadron Collider. By using an ab-initio model for the early stage of the collision combined with Bayesian inference techniques for the first time, we increased the model's predictivity, achieving excellent agreement with measurements used in the model-to-data comparisons, as well as additional measurements not used in the calibration. We found a relatively temperature-independent value for the shear viscosity over entropy density ratio, and showed that a similar description of measurements can be obtained with constant value for $\eta/s$. It is a milestone that the resulting model, with a constant $\eta/s$ and ab-initio initial condition model has eight parameters, approximately \emph{half} the number used in other modern models of heavy-ion collisions, yet achieves a better agreement with data.

We emphasize the importance of such a considerable reduction in the number of parameters used in the multistage model. It is strong support for the overall validity of the model to have an excellent agreement with data using a small number of parameters. This in turn provides support for the constraints on the shear and the bulk viscosity obtained in this work. 

Building upon previous work~\cite{Liyanage:2022byj}, we further showed the benefits of transfer learning methods to study model uncertainties, which we used to account for hydrodynamics-to-transport uncertainty at a fraction of the cost of previous studies. By exploiting this tool, we expect that a much broader range of uncertainties can be explored in the near future, making possible a fuller picture of the transport coefficients of Quantum Chromodynamics.

\acknowledgements We are happy to acknowledge useful exchanges with
members of the JETSCAPE Collaboration. In addition,
we are grateful for useful discussions with S. Bass, D.
Everett, D. Liyanage, S. McDonald, N. Miro-Fortier, and
M. Singh. This work was supported in part by the Natural Sciences and Engineering Research Council of Canada (NSERC), in part by the U.S. Department of Energy Grant no. DE- FG-02-05ER41367, and in part by Vanderbilt University. Computations were performed on the Narval, B\'eluga, Niagara, Graham, and Cedar supercomputers managed by the Digital Research Alliance of Canada and its regional partners, Calcul Qu\'ebec, Compute Ontario, SciNet, SHARCNET, and WestGrid.

\bibliography{Bibliography.bib}
\end{document}